# Connected and Automated Vehicle Distributed Control for On-ramp Merging Scenario: A Virtual Rotation Approach

Tianyi Chen[a], Meng Wang[b], Siyuan Gong[c], Yang Zhou[a,*], Bin Ran[a]

[a] Department of Civil and Environment Engineering, University of Wisconsin-Madison, Madison, WI 53706, USA
[b] Faculty of Civil and Geospatial Engineering, Delft University of Technology, Stevinweg 1, 2628 CN, Delft, the Netherlands
[c] School of Information Engineering, Chang'an University, Xi'an, Shanxi 710064, China
[*] Corresponding Author (Email: zhou295@wisc.edu)

**Abstract**

This study proposes a rotation-based connected automated vehicle (CAV) distributed cooperative control strategy for an on-ramp merging scenario. By assuming the mainline and ramp line are straight, we firstly design a virtual rotation approach that transfers the merging problem to a virtual car following (CF) problem to reduce the complexity and dimension of the cooperative CAVs merging control. Based on this concept, a multiple-predecessor virtual CF model and a unidirectional multi-leader communication topology are developed to determine the longitudinal behavior of each CAV. Specifically, we exploit a distributed feedback and feedforward longitudinal controller in preparation for actively generating gaps for merging CAVs, reducing the voids caused by merging, and ensuring safety and traffic efficiency during the process. To ensure the disturbance attenuation property of this system, practical string stability is mathematically proved for the virtual CF controllers to prohibit the traffic oscillation amplification through the traffic stream. Moreover, as a provision for extending the virtual CF application scenarios of any curvy ramp geometry, we utilize a curvilinear coordinate to model the two-dimensional merging control, and further design a local lateral controller based on an extended linear-quadratic regulator to regulate the position deviation and angular deviation of the lane centerlines. For the purpose of systematically evaluating the control performance of the proposed methods, numerical simulation experiments are conducted. As the results indicate, the proposed controllers can actively reduce the void and meanwhile guarantee the damping of traffic oscillations in the merging control area.

## 1. Introduction

Vehicle merging at highway entrances has received broad attention during the last decades since frequent and nonsmoothed lane-changing (LC) maneuvers significantly impact traffic safety, congestion, and fuel consumption in a merging zone and its ambient areas (McCartt et al., 2004; Papageorgiou et al., 2008; Waard et al., 2009). To mitigate the negative effect of the nonsmoothed merging, most research (Carlson et al., 2010; Sridhar et al., 2008) focuses on traffic flow control and optimization at a macroscopic level that minimizing capacity drop and improve sustainable traffic throughput. One of the most conventional solutions is ramp metering, which restricts the on-ramp vehicles from entering the mainline for conflict reduction and corresponding discharging rate improvement (Carlson et al., 2010b; Cassidy & Rudjanakanoknad, 2005; Papageorgiou et al., 1997). However, due to the microscopic nature of the merging process, the macroscopic traffic flow control methods decrease merging conflicts by dynamically controlling the inflow from ramps to the merging area but cannot eliminate nonsmoothed LC maneuvers. The microscopic analysis reveals that the invasive influence of merging to the mainstream is not negligible.

For example, Laval & Daganzo (2006) claimed that LC vehicles created voids (waste spaces) in the traffic stream, reducing traffic throughput from a macroscopic level. Chen & Ahn (2018) investigated the mechanisms of how spatially distributed LCs interact with capacity-drop at a microscopic level, and the result illustrated that Human-Driven Vehicles (HDVs) created a void in merging that persists downstream due to its lower insertion speed and bounded acceleration, which led to capacity-drop. Additionally, the vehicular interaction across lanes causes disturbances. Due to the instability of HDVs car-following (CF) models, the disturbances will be amplified through vehicle string, leading to frequent stop-and-go maneuvers. The phenomenon is also known as traffic oscillations, significantly impacting the mainstream flow (Ahn & Cassidy, 2007; Ahn et al., 2010; Zheng et al., 2011; Li et al., 2010).

The emerging connected and automated vehicles (CAVs) provides an excellent opportunity to enhance road safety and capacity to resolve the transportation problems caused by humans (Rajamani et al., 2000). Specifically, CAVs bring unprecedented promises to control CAV microscopically by actively dampen traffic oscillations by cooperative adaptive cruise control (Gong & Du, 2018; Wang, 2018; Zhou et al., 2020; Zhou & Ahn, 2019; Montanino et al., 2021; Montanino & Punzo 2021). For the merging control, Schmidt & Posch, (1983) introduced a heuristic approach merging control algorithm for CAVs based on a two-layer controller, the upper-layer and lower-layer controllers. The upper layer determines the vehicle merging sequence, and the lower layer determines the local merging maneuvers of each vehicle. For an upper-level controller, different optimal car sequencing models have been developed. For example, Wang et al., (2007), Ntousakis et al., (2014), and Chen et al., (2020) proposed scheduling methods based on 1) distance from the merging point, 2) traveling time to the merging point, 3) the optimized future car merging sequencing with a predefined objective and constraints. Although these studies schedule the car sequence, a detailed CAV control is needed to execute the optimal sequence, which pertains to the lower-layer controller.

The lower-level controller determines vehicle trajectories to smooth traffic and guarantee merging safety. Based on different levels of cooperation, the prevailing control algorithms can be further categorized as cooperative and non-cooperative control algorithms. The difference between these two algorithms is whether multiple-source information (predecessor vehicles) is used to control the vehicle and the degree of cooperation. For example, Kachroo & Li, (1997) and Lu et al., (2004) presented two non-cooperative single vehicle control algorithms in keeping the safe merging process and smoothing the ambient traffic flow. Specifically, Lu et al., (2004) proposed a 'virtual platoon' concept that projects the CAVs on both the main lane and merging lane to a virtual lane and formulate a 'virtual platoon' before the on-ramp merging point, which transferred the merging problem to a car-following control issue. However, the non-cooperative control algorithm only focused on the individual vehicle optimal decisions, which may induce system-level sub-optimality. The traffic disturbances analysis is ignored even though the 'virtual platoon' concept already enables the stability analysis of the traffic flow.

Conversely, due to the collaborative decision-making and multiple-source information, the cooperative control strategies have greater potentials to further improve system-level behavior and get more attention. Ran et al., (1999) and Davis, (2007) proposed a cooperative control algorithm that created gaps on the mainline large enough for vehicles with CACC function merged without speed reduction by designed CF models. Recently, several advanced control methods have been used in the designing of lower-level control, such as linear (Wang et al., 2019; Xu et al., 2019) and nonlinear controllers (Hu et al., 2021), and model predictive control (MPC). For example, Cao et al., (2015) proposed a cooperative merging path generation method using an MPC scheme to cooperate with the two vehicles on the mainline and ramp, which is closest to the merging point, to accelerate/decelerate smoothly when the merge happens. However, the influence of disturbances of the merging process on the upstream is not theoretically analyzed. Rios-Torres et al., (2015) proposed an optimal vehicle control model for multiple vehicles which targets minimizing the acceleration for safety and fuel consumption cooperatively. Further, Duret et al., (2019) developed a hierarchical control strategy for truck platooning near merging. The lower operational layer applied a MPC for multiple vehicles to actively generate gaps for vehicles to merge according to the merging sequence from the upper tactical layer and guarantee the smoothness of vehicular acceleration and deceleration.

Though the local stability of the longitudinal controller has been analyzed, the stability of the entire merging section has been rarely discussed.

As cited above, most studies focus on ensuring merging safety by actively creating a gap and control the merging vehicles to reduce the void. However, they ignore the cause of traffic oscillations (e.g., caused by the cut-in maneuver and subsequently speed variation) during the process of active generation of gaps, and a corresponding method to dampen the above disturbances during this process. Especially, they lack a mathematical framework and methods to theoretically analyze the system stability for the merging process. Last but not least, the works mentioned above treat the ramp as a straight line, and ignore the lateral movement for vehicles, which limits the width of practical application for different merging section geometric characteristics.

Motivated by the above research gaps, this study firstly designs a virtual rotation approach that transfers the merging problem of straight ramp lines to a CF problem by the concept proposed by Lu et al., (2004). This concept has been extended in Xu et al., (2018) at an unsignalized intersection which further inspired us to systematically design the virtual rotation approach. Thus, the rotation process, which serves as an upper-level controller, uses a predetermine merge point as a reference to calculate the relative spacing for vehicle and then determines the virtual car following sequences of vehicles in a predefined merging control area. A lower-level cooperative distributed control strategy is proposed to control vehicles' trajectories with a specifically designed unidirectional multi-leader communication topology (MLT). In detail, this approach exploits a weighted multiple predecessor information-based linear feedback and feedforward controller to regulate the virtual platoon spacing and speed differences. The string stability for the proposed controller is further mathematically proven. To generalize the application scenario, this study extends the framework to a two-dimensional case by considering CAVs lateral movement. Specifically, a curvilinear coordinate is utilized to describe the ramp with arbitrary shape. A local-based lateral controller is developed to control the lateral motion of CAVs, while still satisfying the concept, design, and the stability proof for the straight ramp line scenario mentioned above.

The rest of this paper is organized as follows. **Section 2** briefly introduces the problem statement. **Section 3** utilized a virtual rotation concept to reformulate the problem as a 'virtual' CF problem assuming the ramp is straight. Based on that, we propose a communication topology and formulate the cooperative control model on the shared virtual lane. For rigor, **Section 4** proves the string stability criterion for the 'virtual' CF problem. In **Section 5**, we relax the assumption of the straightness of the ramp and develop a local lateral controller by extending the ramp scenario to a two-dimensional path coordinate. **Section 6** provides numerical simulation experiments to show the effectiveness of our algorithm. Last, we give the conclusion and point out future works in **Section 7**.

## 2. Control objectives and assumptions

In this study, a cooperative merging control strategy is developed for CAVs at an on-ramp area which actively reduces the void and meanwhile guarantees the stability of the merging area. Specifically, given the initial position and velocity of each CAVs and the road geometric information, the following steps are conducted to fulfill the control target mentioned above: 1) determine the virtual car sequence based on the known position and velocity of each vehicle in a centralized manner (e.g., per 1 sec); 2) design the communication topology for each vehicle according to the car sequencing order; 3) determine the acceleration portfolio for vehicles in the merging control area (e.g., 0.01 sec); 4) determine the vehicle steering angle portfolio (e.g., 0.01 sec) in the merging control area as an extension based on step 3.

The design is restricted to a pure CAV environment. And the merging geometric consists of one lane on each road, as shown in **FIG. 1**. Ahead of detailed modeling, the following assumptions have been adopted:

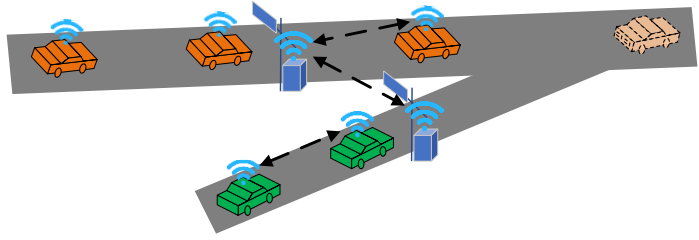

**FIG 1.** Schematic diagram of the proposed merging at on-ramp control area scenario

- The infrastructure has "intelligence" (Ding et al., 2019; Ran et al., 2020) that can sense vehicle information and road geometrics information over a predefined merge control area.
- The CAVs are fully automated (SAE, 2016), which means they can be controlled by themselves. Furthermore, they can communicate with each other and with infrastructure.
- Merge control area is ample enough (e.g., the control begin area is at least 250 meters away from the merging point) to apply the control strategy before vehicles merge.
- The second-order vehicle dynamics are considered for this study.
- The communication delays are negligible due to the increasing maturity of 5G communication technologies (Akpakwu et al., 2017).
- The control algorithm obeys the "First in first out" (FIFO) scheme.

## 3. System modeling

This section describes the design and formulation of the proposed cooperative merging control strategy for a simple scenario where the mainline and ramp line is straight. A virtual rotation strategy has been introduced to reduce modeling complexity. Specifically, this study rotates CAVs on mainline ($Y_1$-axis) and CAVs on on-ramp ($Y_2$-axis) to a shared straight axis $Z$, as shown in Fig.2. A virtual CF sequence is obtained by the rotation strategy and FIFO scheme. To leverage the V2V communication among multi-vehicles, a unidirectional multi-leader communication topology is proposed in this research. According to the rules of sequence and topology, a multi-predecessor linear feedback and feedforward controller has been exploited to alleviate the disturbances on the gap generation and merging procedure.

### 3.1. Virtual rotation and virtual car following sequencing

To distinguish the CAVs, we define two groups of sets for CAVs on mainline (i.e., $I_M = \{1_M, 2_M, \dots m_M\}$) and on-ramp (i.e., $I_R = \{1_R, 2_R, \dots r_R\}$), respectively. Each group of sets contains a vehicle sequence set, a vehicle position set, and a vehicle velocity set, which is denoted as $X_M = \{x_1, x_2, \dots, x_m\}, X_M \in \mathbb{R}^m$, $V_M = \{v_1, v_2, \dots, v_m\}, V_M \in \mathbb{R}^m$, $X_R = \{x_1, x_2, \dots, x_r\}, X_R \in \mathbb{R}^r$, $V_R = \{v_1, v_2, \dots, v_r\}, V_R \in \mathbb{R}^r$.

Based on the vehicles' state information, we define the concept of 'virtual rotation' as the process determining the car-following sequence of merging, which is equivalent to finding the optimal car following sequence for vehicles in a virtual axle based on a predefined law. The importance of the maneuver lies in reducing modeling complexity and convert the merging problem to a car following problem, which makes the string stability analysis available. One of the most straightforward strategies is sorting the car following sequence based on the relative distance to the merging point. Precisely, we rotate the on-ramp $y_2$-axis to the mainline $y_1$-axis and construct a virtual $Z$-axis, while maintaining the physical distance to the merge point as illustrated by **FIG 2**. Based on that, we can have a union of vehicles index, relative position, and speed set on the virtual axis $Z$, as below:

$$X_Z = X_M \cup X_R, \tag{1a}$$

$$V_Z = V_M \cup V_R, \tag{1b}$$
$$I_Z = I_M \cup I_R. \tag{1c}$$

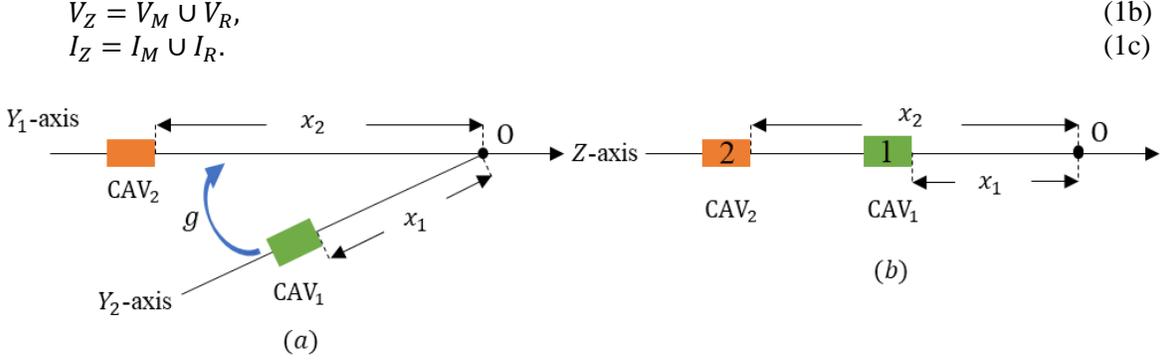

**FIG 2.** An example of the proposed virtual rotation approach in (a) $Y_1$-axis, $Y_2$-axis, (b) virtual Z-axis

Correspondingly, based on the CAVs' index and positions on the Z-axis, we can quickly get the virtual car-following sequences on Z-axis by a function $g$ which sorts $X_Z$ in a monotonically descending order:

$$(\tilde{X}_Z, \tilde{V}_Z, \tilde{I}_Z, \tilde{F}_Z) = g(X_Z, V_Z, I_Z), \tag{2}$$

where $\tilde{X}_Z, \tilde{V}_Z, \tilde{I}_Z, \tilde{F}_Z$ represent the relative position set, speed set, car following sequence set, the ramp/mainline indicator set on Z-axis after sorting. $\tilde{I}_Z$ is organized in an ascending order $\tilde{I}_Z = \{1,2,\ldots,N_{total}\}$, $N_{total} = M + R$, $\tilde{F}_Z = \{f_{Z,1}, \ldots f_{Z,N_{total}}\}$. Specifically, for $i \in \tilde{I}_Z$, we let $f_{Z,i} = 1$, if the $i^{th}$ vehicle on the virtual axis Z is actually on the mainline, and $f_{Z,i} = 0$ vice versa. To be noticed that, if two vehicles have the same distance to the merging point ($\tilde{I}_Z$ has two same values), $g$ would sequence them based on their speeds ($V_Z$) since the probability of two vehicles keeping the same distance to the merging point is zero when the distance is a continuous variable. As an upper-level controller, $g$ is updated when the vehicle enters the control area at the very first time or updated every 5 sec. Though this virtual rotation maneuver may result in violation of safety in the virtual axle, the vehicles that have safety violations are actually at different lanes. Moreover, since the rotation process sequences all vehicles in the merging area, the mainline and on-ramp vehicles will only have equal or large desired spacing. Note that, $g$ can be any function satisfying FIFO, such as Chen et al., (2020) can be applied to our framework. We just utilized a simple strategy as Eq. (2) since this section is not the main focus of this paper.

### 3.2. Communication topology design

According to the above virtual car-following sequence, we specifically design a unidirectional multi-leader communication topology (MLT) to leverage the V2V communication among multi-vehicles and prevent collisions between adjacent cars on both mainline and ramp entrances. The communication topology will further facilitate the strict string stability for the vehicles on the mainline and ramp, respectively, later given in Section 4.

The MLT on the virtual Z-axis is modeled with a directed graph $\mathcal{G}(\tilde{I}_Z, \mathcal{E}, \mathcal{A})$, where $\tilde{I}_Z$ represents the vertices of the graph, whose number of vertices is $N_{total}$. $\mathcal{E} \subseteq \tilde{I}_Z \times \tilde{I}_Z$ is the set of edges representing the car following connections between each pair of following vehicles. The adjacency matrix $\mathcal{A} = [a_{i,j}], \forall i,j \in \tilde{I}_Z$, represents the communication connections. We specifically let a complementary communication set as $\mathcal{E}_c$ connecting vertices $\forall i,j \in \tilde{I}_Z$, and $i < j$, defined as: $\mathcal{E}_c = \cup\{edge\ (i,j)\}$ if $f_{Z,i} = f_{Z,j}$ & $i < k < j$, $f_{Z,k} = f_{Z,i}$ or $f_{Z,i} \neq f_{Z,j}$ & $i < k < j$, $f_{Z,k} = f_{Z,j}$. The edge set $\mathcal{E}$ is defined as:

$$\mathcal{E} = \{\tilde{I}_Z \times \tilde{I}_Z\} \setminus \mathcal{E}_c \tag{3}$$

And the $\mathcal{A}$ can be defined as:

$$\begin{cases} 1, if\{\mathcal{E}_{i,j} \in \mathcal{E}\}, \\ 0, otherwise, \end{cases} i,j \in N_{total}, \qquad (4)$$

where $\mathcal{E}_{i,j} \in \mathcal{E}$ indicates that vehicle $j$ can receive the information of vehicle $i$.

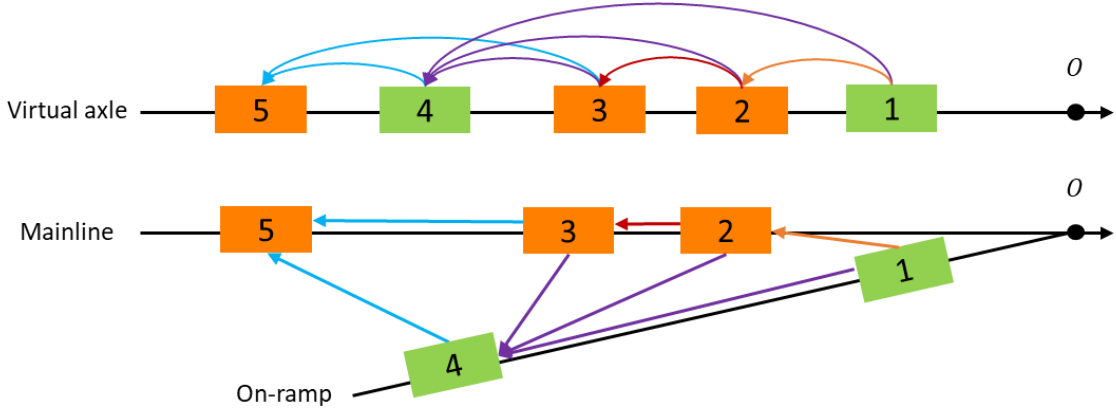

**FIG 3.** An example of MLT in both virtual and real-world scene

To clearly understand the definition of Eq. (3), we summarize it as the thumb of rules below: 1) for vehicle $i$ and predecessor $j$ with the same ramp/mainline indicator if there is a vehicle $k$ between $i$ and $j$ that has the same indicator as vehicle $i$, no communication is allowed between $i$ and $j$; 2) for vehicle $i$ and predecessor $j$ with the different indicator, if there is a vehicle $k$ between $i$ and $j$ that has the same indicator as predecessor $j$, no communication is allowed between $i$ and $j$. Other than the two above scenarios, CAVs are all unidirectional interconnected. An intuitive example has been provided in **FIG. 3** to illustrate how CAVs communicate in both virtual and real-world scenes. This example contains 5 CAVs, 3 on the mainline, 2 on the on-ramp, and the virtual car-following sequence has been determined as $\tilde{F}_Z = \{1,0,0,1,0\}$. Based on the MLT rules, the fifth vehicle (the last mainline vehicle passing the merging point) can only communicate with its two predecessors. No communication is allowed with the second and first CAV since the third CAV already has the same mainline indicator as the fifth one. Similarly, the fourth CAV can receive information from its three predecessors until the first vehicle since the first vehicle has the same on-ramp indicator.

### 3.3. Longitudinal control structure

The merging problem can be converted to a virtual CF control problem by virtual rotation and virtual car following sequencing. Specifically, the control objective is to coordinate CAVs on-ramp and mainline to actively generate the gap for vehicles to merge, reduce the voids, and meanwhile reduce traffic oscillations during the whole process. Based on the rotation scheme developed above, we design a linear feedback and feedforward controller based on Wang et al., (2020) for vehicles within each subset.

The schematic of the feedback and feedforward controller for vehicle $i$ is illustrated in **FIG. 4**. $U_i$ represents the control command, which consists of two feedback control components $U_{b,i}$: 1) the spacing error $E_i$, 2) the velocity difference $\Delta V_i$, and one feedforward control term $U_{f,i-k}$ from the acceleration rates $\ddot{x}_{i-k}$ of every predecessors from vehicle $i-1$ to $i-k$. $G_i$ and $M_i$ are the ideal longitudinal vehicle dynamics. $X_i$ and $V_i$ represents the position and velocity output of vehicle $i$. $H_i$ denotes the constant time headway (CTH) spacing policy. $\omega_{e,i}$ and $\omega_{v,i}$ are the equilibrium spacing coefficient and equilibrium velocity coefficient. $\alpha_{f,k}$ and $\alpha_{b,k}$ are the weighting coefficients for acceleration feedback and feedforward information. $\gamma_{b,k}$ is the weighting coefficient for velocity feedback information. These three coefficients represent the relative importance of the information from $k$ predecessors to vehicle $i$ and are different from each case which are discussed in the following sessions.

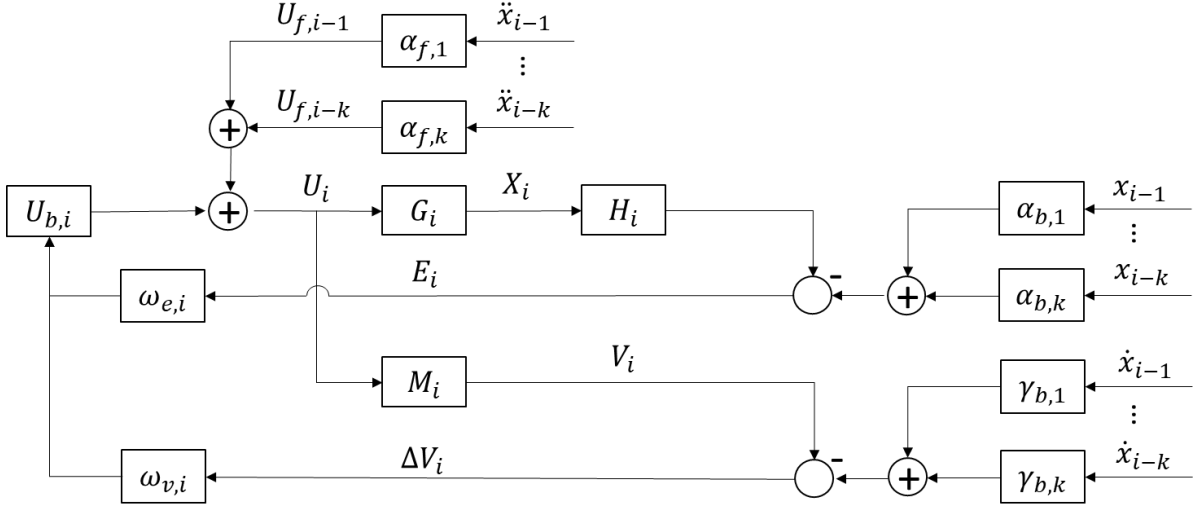

**FIG 4.** Schematic of the feedback and feedforward controller

This study considers an idealized longitudinal vehicle dynamics model that ignores realist factors such as air drag and actuator delay. For the lower-level controller, we assume it can address vehicle internal dynamics so that vehicle can respond to acceleration and velocity commands without any delays. This assumption is widely used in the literatures, e.g., (Wang et al., 2020; Zhou et al., 2017). The linearized state-space representation of the idealized longitudinal vehicle dynamics can be represented as:

$$\begin{cases} \dot{x}_i(t) = v_i(t), \\ \dot{v}_i(t) = u_i(t), \end{cases} \tag{5}$$

where $x_i(t)$, $v_i(t)$ and $u_i(t)$ are the absolute position, velocity, and acceleration of vehicle $i$ at time t.

To simplify the further stability analysis, the modeling and analysis are performed in the Laplace domain. The idealized longitudinal vehicle dynamics model in the Laplace domain can be described by using transfer functions as:

$$G_i(s) = X_i(s)U_i(s)^{-1} = s^{-2}, \tag{6a}$$
$$M_i(s) = V_i(s)U_i(s)^{-1} = s^{-1}, \tag{6b}$$

where the input $U_i(s)$ denotes the acceleration of vehicle $i$ and the output $X_i(s)$ and $V_i(s)$ denotes the absolute position and velocity of vehicle $i$ in the Laplace domain, respectively.

To achieve more efficient damping oscillations, we obtain the desired relative distance between vehicle $i$ and its $N_i$ predecessors, whose communications are active, using the CTH policy as follows:

$$d_{i,k}(t) = k[L + \tau \dot{x}_i(t)], \tag{7}$$

where $d_{i,k}(t)$ is the desired relative distance between vehicle $i$ and vehicle $k$, and $\tau$ is the desired time gap for vehicle $i$. $L$ is the constant standstill distance (including vehicle length) between the two adjacent vehicles, $\dot{x}_i(t)$ is the velocity of vehicle $i$.

The convex combination of spacing errors is as follows:

$$e_i(t) = \sum_{k=1}^{N_i} \alpha_{b,k}\{[(x_{i-k}(t) - x_i(t)] - d_{i,k}(t)\}, \tag{8}$$

s.t.

$$\sum_{k=1}^{N} \alpha_{b,k} = 1, \tag{9}$$

where $\alpha_{b,k}$ is the weighting coefficients for position feedback information.

Substituting Eq. (7) into Eq. (8), the weighted spacing error is

$$e_i(t) = \sum_{k=1}^{N_i} \alpha_{b,k}\{[(x_{i-k}(t) - x_i(t)] - k[L + \tau \dot{x}_i(t)]\}. \tag{10}$$

Taking the Laplace transformation of Eq. (10), the spacing error can be expressed equivalently as:

$$\begin{aligned} E_i(s) &= \sum_{k=1}^{N_i} \alpha_{b,k}\{[(X_{i-k}(s) - X_i(s)] - k\tau s X_i(s)\} \\ &= \sum_{k=1}^{N_i} \alpha_{b,k} X_{i-k}(s) - X_i(s)\left(1 + \sum_{k=1}^{N_i} \alpha_{b,k} k\tau s\right) \\ &= \sum_{k=1}^{N_i} \alpha_{b,k} X_{i-k}(s) - H_i(s) X_i(s), \end{aligned} \tag{11}$$

where $H_i(s)$ is the CTH spacing policy in the frequency domain, given by:

$$H_i(s) = 1 + \sum_{k=1}^{N_i} \alpha_{b,k} * k * \tau * s. \tag{12}$$

The equilibrium velocity of vehicle $i$ with its $N_i$ predecessors can be represented as:

$$v_{i,e}(t) = \sum_{k=1}^{N_i} \gamma_{b,k} v_{i-k}(t), \tag{13}$$

s.t.

$$\sum_{k=1}^{N_i} \gamma_{b,k} = 1, \tag{14}$$

where $\gamma_{b,k}$ is the weighting coefficient for velocity feedback information. Then, the deviation from equilibrium velocity of vehicle $i$ can be expressed as:

$$\Delta v_{i,e}(t) = v_i(t) - v_{i,e}(t). \tag{15}$$

Taking the Laplace transformation of Eqs. (13) and (15), we have:

$$V_{i,e}(s) = \sum_{k=1}^{N_i} \gamma_{b,k} V_{i-k}(s). \tag{16}$$

$$\Delta V_{i,e}(s) = V_i(s) - \sum_{k=1}^{N_i} \gamma_{b,k} V_{i-k}(s). \tag{17}$$

The control command consists of a feedback term (including spacing feedback and velocity feedback) and a set of feedforward terms, which is shown as follows:

$$U_i(s) = U_{b,i}(S) + \sum_{k=1}^{N_i} U_{f,i-k}(S), \tag{18}$$

where the feedback term $U_{b,i}(S)$ uses spacing and velocity errors to stabilize the closed-loop system while the feedforward term $U_{f,i-k}(S)$ uses acceleration rates from the $N_i$ predecessors to minimize the spacing error.

The feedback term $U_{b,i}(S)$ and the corresponding feedback controller are defined as:

$$U_{b,i}(S) = \omega_{e,i} E_i(s) + \omega_{v,i} \Delta V_{i,e}(s), \tag{19}$$

where $E_i(s)$ is the spacing error in the Laplace domain in Eq. (11). $\omega_{e,i}$ and $\omega_{v,i}$ are the feedback gains for deviation from equilibrium spacing and equilibrium speed, respectively, where larger feedback gains will make vehicle react more intensively.

The feedforward terms $U_{f,i-k}(S)$ indicate that the acceleration rates of vehicle $i - k$ are defined as:

$$U_{f,i-k}(S) = \alpha_{f,k} s^2 X_{i-k}(s). \tag{20}$$

The overall control command is obtained by summing up Eqs. (18), (19) and (20). Through inverse Laplace transformation, the expression for the control command is:

$$U_i(t) = \omega_{e,i} e_i(t) + \omega_{v,i} \Delta v_{i,e}(t) + \sum_{k=1}^{N_i} \alpha_{f,k} \ddot{x}_{i-k}(t). \tag{21}$$

## 4. String stability analysis

In this section, we analyze the string stability for the longitudinal controller with the MLT mentioned in Section 3.2. Specifically, we mathematically define the property of string stability and derive sufficient conditions by the analysis method. Since the velocity of the vehicle $i$ is affected by its multiple predecessors, we consider the following definition of $L_2$ norm string stability based on (Darbha et al., 2017):

**Definition 1.** A CAV platoon is (practically) $L_2$ norm string stable if and only if:

$$\|v_i(t)\|_2^2 \leq \frac{1}{N_i} \sum_{k=1}^{N_i} \|v_{i-k}(t)\|_2^2, \tag{22}$$

where $v_i(t)$ is the absolute velocity of vehicle $i$ on the virtual Z-axis. $\|v_i(t)\|_2^2 = \int_{-\infty}^{+\infty} |v_i(t)|^2 \, dt$ is the square of the $L_2$ norm of $v_i$. According to Eq. (22), it requires that the square of the $L_2$ absolute velocity of vehicle $i$ is attenuated in the sense that it is less than the average of the squares of its predecessors' $L_2$

absolute velocities. Note that, in the following study, we treat the square of the $L_2$ absolute velocity as energy for easy understanding.

As widely known, the frequency domain facilites the behavior analysis of the system string stability. Hence, we conduct the Fourier transformation on the system by Eqs. (11), (17), and (21). The result is shown in Eq. (23).

$$s^2 X_i(s) = \omega_{e,i} \left[ \sum_{k=1}^{N_i} \alpha_{b,k} X_{i-k}(s) - H_i(s) X_i(s) \right] + \omega_{v,i} s \left[ X_i(s) - \sum_{k=1}^{N_i} \gamma_{b,k} X_{i-k}(s) \right] + s^2 \sum_{k=1}^{N_i} \alpha_{f,k} X_{i-k}(s). \tag{23}$$

To simplify the further analysis, we define $Q_k(s) = \frac{P_k(s)}{J_k(s)}$, where $J_k(s) = s^2 + \omega_{e,i} H_i(s) - s\omega_{v,i}$, and $P_k(s) = \omega_{e,i}\alpha_{b,k} - s\omega_{v,i}\gamma_{b,k} + s^2 \alpha_{f,k}$. Then, Eq. (23) can be represented as:

$$X_i(s) = \sum_{k=1}^{N_i} Q_k(s) X_{i-k}(s). \tag{24}$$

By dividing $s$ on both sides of Eq. (24), we can have:

$$V_i(s) = \sum_{k=1}^{N_i} Q_k(s) V_{i-k}(s). \tag{25}$$

Based on Definition 1, and Eq. (25), the Proposition 1 can be got as below:

**Proposition 1.** If $\|Q_k(j\omega)\|_\infty \leq \frac{1}{N_i}$, the CAV platoon is string stable on the virtual $Z$-axis.

**Proof.** According to the Parseval's theorem (Arfken et al., 2013), the energy in time and frequency domain are equal, and then we have:

$$\|v_i(t)\|_2^2 = \int_{-\infty}^{+\infty} |v_i(t)|^2 dt = \int_{-\infty}^{+\infty} |V_i(j\omega)|^2 d\omega = \int_{-\infty}^{+\infty} \left| \sum_{k=1}^{N_i} (Q_k(j\omega) V_{i-k}(j\omega)) \right|^2 d\omega. \tag{26}$$

According to the Cauchy-Schwarz inequality (Steele, 2004), we can get,

$$\|v_i(t)\|_2^2 \leq \int_{-\infty}^{+\infty} \left( N_i \sum_{k=1}^{N_i} (V_{i-k}^T(j\omega) Q_k^T(j\omega) Q_k(j\omega) V_{i-k}(j\omega)) \right) d\omega$$

$$\leq N_i \sum_{k=1}^{N_i} \left( (\sup_\omega |Q_k(j\omega)|)^2 \cdot \int_{-\infty}^{+\infty} V_{i-k}^T(j\omega) V_{i-k}(j\omega) d\omega \right)$$

$$= N_i \sum_{k=1}^{N_i} (\|Q_k(j\omega)\|_\infty^2 \cdot \|v_{i-k}(t)\|_2^2). \tag{27}$$

To satisfy the definition 1, by substituting Eq. (27) into (22), we can have:

$$\|Q_k(j\omega)\|_\infty \leq \frac{1}{N_i}, \forall\ 1 \leq k \leq N_i. \tag{28}$$

Next, we bring the Eq. (28) proposed by proposition 1 into the two cases we defined below:

1. *The equally weighted case:*

We firstly begin with Case 1 where weighted coefficients are equal to $\frac{1}{N_i}$ similar to Bian et al., (2019), which are revealed below in Eq. (29). The exact identical coefficients are for dimensionality reduction, which neat the parameters and control algorithm. And the assignment of equal weight demonstrates that we treat every coefficient under each $k$ is equally important.

$$\alpha_{b,k} = \alpha_{f,k} = \gamma_{b,k} = \frac{1}{N_i}. \tag{29}$$

**Proposition 2.** The CAV platoon is string stable on the virtual Z-axis, if the following inequality equation is satisfied:

$$\left(\omega_{e,i}\tau\frac{(1+N_i)}{4}\right) - \omega_{v,i} \geq 0. \tag{30}$$

**Proof.** By substituting Eq. (29) into Eq. (28), we can get:

$$\|Q_k(j\omega)\|_\infty = \left\|\frac{P_k(j\omega)}{J_k(j\omega)}\right\|_\infty = \left\|\frac{\omega_{e,i}\alpha_{b,k} - j\omega\omega_{v,i}\gamma_{b,k} - \omega^2\alpha_{f,k}}{-\omega^2 + \omega_{e,i}(1 + \sum_{k=1}^{N_i}\alpha_{b,k} * k\tau j\omega) - j\omega\omega_{v,i}}\right\|_\infty$$

$$= \left\|\frac{\frac{1}{N_i}\omega_{e,i} - \frac{1}{N_i}\omega^2 - \frac{1}{N_i}\omega_{v,i}j\omega}{-\omega^2 + \omega_{e,i} + \left(\left(\omega_{e,i}\tau\frac{(1+N_i)}{2}\right)\omega - \omega\omega_{v,i}\right)j}\right\|_\infty \leq \frac{1}{N_i}. \tag{31}$$

By multiply $N_i$ on both sides of Eq. (31) and simplified to $\left\|\frac{A+Bj}{C+Dj}\right\|_\infty \leq 1$, we have:

$$A = \omega_{e,i} - \omega^2, \tag{32a}$$
$$B = -\omega\omega_{v,i}, \tag{32b}$$
$$C = -\omega^2 + \omega_{e,i}, \tag{32c}$$
$$D = \left(\omega_{e,i}\tau\frac{(1+N_i)}{2}\right)\omega - \omega\omega_{v,i}. \tag{32d}$$

Then, the Eqs. (32a)–(32d) can be simplified as $A^2 + B^2 \leq C^2 + D^2$, which is shown below:

$$\omega_{e,i}^2 + \omega^4 - 2\omega^2\omega_{e,i} + \omega^2\omega_{v,i}^2$$
$$\leq \left(\omega_{e,i}^2\tau^2\frac{(1+N_i)^2}{4}\right)\omega^2 + \omega^2\omega_{v,i}^2 - \left(\omega_{e,i}\tau(1+N_i)\right)\omega^2\omega_{v,i}$$
$$+ \omega^4 + \omega_{e,i}^2 - 2\omega^2\omega_{e,i} \Rightarrow \left(\omega_{e,i}\tau\frac{(1+N_i)}{4}\right) - \omega_{v,i} \geq 0. \tag{33}$$

In a special case scenario, when $N_i = 1$, Eq. (33) can be expressed as:

$$\frac{\omega_{e,i}\tau}{2} - \omega_{v,i} \geq 0. \tag{34}$$

*2. The non-equally weighted case:*

We extend case 1 to a more general and realistic case where the weighted coefficients are assigned by a function of $k$ and $N_i$ which shows below in Eq. (35). The assignment function is to show that the closer to $CAV_i$, the greater the weighted coefficient of the CAV and the greater the impact on $CAV_i$.

$$\alpha_{b,k} = \alpha_{f,k} = \gamma_{b,k} = \begin{cases} \dfrac{1}{2^k}, & 1 \leq k \leq N_i - 1 \\ \dfrac{1}{2^{N_i-1}}, & k = N_i \end{cases}. \tag{35}$$

**Theorem 1** (Darbha et al., 2019). The string stability for the system with the form as Eq. (25) is string stable if the following inequality is satisfied:

$$\sum_{k=1}^{N_i} \|Q_k(j\omega)\|_\infty \leq 1. \tag{36}$$

**Proposition 3.** The CAV platoon is string stable on the virtual Z-axis, if the following inequality equation are satisfied:

$$\omega_{e,i}\tau\theta - 2\omega_{v,i} \geq 0. \tag{37}$$

**Proof.** By substituting Eq. (35) into (36), we can have:

$$\sum_{k=1}^{N_i} \|Q_k(j\omega)\|_\infty = \left\|\frac{\omega_{e,i}\alpha_{b,k} - j\omega\omega_{v,i}\gamma_{b,k} - \omega^2\alpha_{f,k}}{-\omega^2 + \omega_{e,i}(1 + \sum_{k=1}^{N_i}\alpha_{b,k}*k\tau j\omega) - j\omega\omega_{v,i}}\right\|_\infty \leq \sum_{k=1}^{N}\alpha_{b,k}. \tag{38}$$

By dividing $\sum_{k=1}^{N_i}\alpha_{b,k}$ on both hand side of Eq. (38) and simplified to $\left\|\frac{A+Bj}{C+Dj}\right\|_\infty \leq 1$, we have:

$$A = \omega_{e,i} - \omega^2, \tag{39a}$$
$$B = -\omega\omega_{v,i}, \tag{39b}$$
$$C = -\omega^2 + \omega_{e,i}, \tag{39c}$$
$$D = (\omega_{e,i}\tau\theta)\omega - \omega\omega_{v,i}, \tag{39d}$$
$$\theta = \sum_{k=1}^{N_i}\alpha_{b,k}*k. \tag{39f}$$

Then, the Eqs. (39a)–(39f) can be simplified as $A^2 + B^2 \leq C^2 + D^2$, which is shown below:

$$\begin{aligned}\omega_{e,i}^2 + \omega^4 &- 2\omega^2\omega_{e,i} + \omega^2\omega_{v,i}^2 \\ &\leq (\omega_{e,i}^2\tau^2\theta^2)\omega^2 + \omega^2\omega_{v,i}^2 - 2\omega^2\omega_{e,i}\omega_{v,i}\tau\theta + \omega^4 + \omega_{e,i}^2 \\ &- 2\omega^2\omega_{e,i} \Rightarrow \omega_{e,i}\tau\theta - 2\omega_{v,i} \geq 0.\end{aligned} \tag{40}$$

**Remark 1.** According to Wang et al., (2020), we consider the worst-case scenario where the $\frac{V_{i-k}(s)}{V_{i-N}(s)}, \forall\ 1 \leq k < N$ is equal to one (which means the traffic oscillation is neither amplified nor dampened), the transfer function in Eq. (25) becomes:

$$V_i(s) = V_{i-N}(s) \sum_{k=1}^{N_i} Q_k(s). \tag{41}$$

By Cauchy-Schwarz inequality, we have:

$$\|V_i(s)\|_2 \leq \left\|\sum_{k=1}^{N_i} Q_k(s)\right\|_\infty \|V_{i-N}(s)\|_2. \tag{42}$$

To satisfy the head-to-tail string stability (Wang, 2018), $\frac{\|V_i(s)\|_2}{\|V_{i-N}(s)\|_2} \leq 1$, which gives the sufficient condition as $\left\|\sum_{k=1}^{N_i} Q_k(s)\right\|_\infty \leq 1$ based on Eq. (42). Further, since

$$\sum_{k=1}^{N_i} \|Q_k(j\omega)\|_\infty \leq \left\|\sum_{k=1}^{N_i} Q_k(s)\right\|_\infty \leq 1, \tag{43}$$

we can find that practical string stability conditions given by Eq. (28) and Eq. (36) is more stringent, and hence the head-to-tail string stability can be naturally satisfied. Then, by combining the head-to-tail string stability and the MLT we designed in Section 3, we can further conclude that the vehicles on the mainline and on-ramp before the merging point are strictly string stable.

## 5. Two-dimension coordinate extension

To increase the application on all road geometries, we extend the ideal merging scenario whose ramp is a straight line to a general scenario, a two-dimensional case. A local lateral controller is designed to regulate the vehicles' deviation from the lane centerline based on the proposed longitudinal control. We firstly describe the vehicle state in a curvilinear coordinates model and illustrate the vehicle lateral dynamics. The local linear feedback and feedforward controller are designed based on an extended linear quadratic regulator (ELQR). Note that to make the system more realistic, both holonomic and nonholonomic vehicle lateral movement system has been considered in this study.

### 5.1. Curvilinear coordinates modelling

This paper uses a curvilinear coordinates model rather than a Cartesian coordinate model to model the vehicle kinematics of the lateral movements. Although the Cartesian coordinate model can easily express the distance between the cars and the merging point, the Curvilinear coordinate model can represent the vehicle trajectory (including steering angle, position, and distance) more accurately and conveniently to the centerline of the curved on-ramp. Since both coordinate models can be invertible transformed by each other, the curvilinear coordinates model given by **FIG. 5.**, can be represented as $[p_x, p_y, \theta]^S \in R^3$: where its position is $[p_x, p_y]$, and $\theta$ is the orientation in the global frame. Then, the vehicle state $M(t)$ can be represented in a curvilinear coordinate $[x_i(t), r_i(t), \Delta\theta_i(t)]^T$, where the vehicle state contains the path length (curvilinear abscissa) at the closest point, $x_i(t)$, lateral deviation $r_i(t)$, the angular deviation $\Delta\theta_i(t)$ which can be expressed as:

$$M_i(t) = \begin{bmatrix} x_i(t) \\ r_i(t) \\ \Delta\theta_i(t) \end{bmatrix}, \quad (44)$$

where $r_i(t)$ is the orthogonal distance from the center of the CAV to the closest point on the on-ramp $y_2$-axis, and $\Delta\theta_i(t) = \theta_i(t) - \theta_{des(i)}(t)$, $\theta_i(t)$ is the angle between the CAV heading direction and the $y_1$-axis, $\theta_{des(i)}(t)$ is the angle between the tangent of the road centerline and the $y_1$-axis in the global frame. An example of Two CAVs under our scenario in Curvilinear coordinate system is shown in **FIG. 5**.

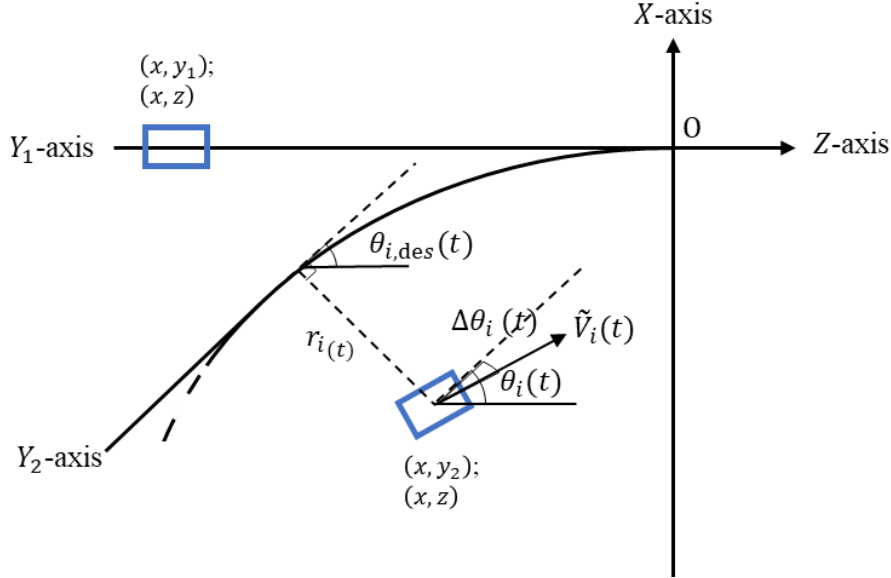

**FIG 5.** Vehicle Dynamics in Curvilinear Coordinates Model

*5.2.  Holonomic vehicle lateral dynamics*

Then, the vehicle dynamics can be described by the state space as below:

$$\frac{dM_i(t)}{dt} = \frac{d}{dt}\begin{bmatrix} x_i(t) \\ r_i(t) \\ \Delta\theta_i(t) \end{bmatrix} = \begin{bmatrix} \tilde{v}_i(t) * \cos(\Delta\theta_i(t)) \\ \tilde{v}_i(t) * \sin(\Delta\theta_i(t)) \\ \mu_i(t) - \dfrac{\tilde{v}_i(t) * \cos(\Delta\theta_i(t))}{R_i(t)} \end{bmatrix}, \quad (45)$$

where $\tilde{v}_i(t)$ is the velocity of the CAV in the curvilinear coordinate. To make the structure and stability given by the longitudinal controller mentioned in Section 3 and 4 still hold, we can simply let $v_i(t) = \tilde{v}_i(t) * \cos(\Delta\theta_i(t))$, and the string stability refers to a projected practically string stability by $\cos(\Delta\theta_i(t))$. $\mu_i(t)$ is the angular velocity and $R_i(t)$ is the radius of the curvature for CAV $i$ at time $t$. Note that, $x_i(t)$ is more related to the system level stability (string stability in our case), while the lateral movement is more related to the individual vehicle stability (usually known as local stability). Hence, for this extension of lateral control, we only need to consider the components of lateral deviation function $dr_i(t)/dt$ and angular deviation function $d\Delta\theta_i(t)/dt$ in Eq. (45). Therefore, we can define the lateral state $M_{i,l}(t)$ as $[r_i(t), \Delta\theta_i(t)]^T$, and the state space function for the lateral vehicle dynamics becomes:

$$\frac{dM_{i,l}(t)}{dt} = \frac{d}{dt}\begin{bmatrix} r_i(t) \\ \Delta\theta_i(t) \end{bmatrix} = \begin{bmatrix} \tilde{v}_i(t) * \sin(\Delta\theta_i(t)) \\ \mu_i(t) - \dfrac{\tilde{v}_i(t) * \cos(\Delta\theta_i(t))}{R_i(t)} \end{bmatrix}. \quad (46)$$

However, the function is still complex because of nonlinearities caused by sine/cosine function. Luckily, the angle $\Delta\theta_i(t)$ is usually small enough (e.g., $\Delta\theta_i(t) < 14°$, which also suggests $v_i(t) \approx \tilde{v}_i(t)$) that satisfies the small-angle approximation to simplify vehicle lateral dynamic which given as follow:

$$\frac{dM_{i,l}(t)}{dt} = \frac{d}{dt}\begin{bmatrix} r_i(t) \\ \Delta\theta_i(t) \end{bmatrix} = \begin{bmatrix} \tilde{v}_i(t) * \Delta\theta_i(t) \\ \mu_i(t) - \frac{\tilde{v}_i(t)}{R_i(t)} \end{bmatrix}. \tag{47}$$

By defining $\mu_i(t)$ as the system control input, Eq. (47) can be expressed as a linear time variant system (LTV) as:

$$\frac{dM_{i,l}(t)}{dt} = A_i(t) * M_{i,l}(t) + B_i(t)\mu_i(t) + D_i(t), \tag{48}$$

where $A_i(t) = \begin{bmatrix} 0 & \tilde{v}_i(t) \\ 0 & 0 \end{bmatrix}$, $B_i(t) = \begin{bmatrix} 0 \\ 1 \end{bmatrix}$, $D_i(t) = \begin{bmatrix} 0 \\ -\frac{\tilde{v}_i(t)}{R_i(t)} \end{bmatrix}$.

To determine the $\mu_i(t)$, we apply a widely applied ELQR by Singh & Pal, (2017) at each time point $t$ in a rolling horizon manner according to the most recent measurement $\tilde{v}_i(t)$, which can be represented as:

$$\min J_i(M_{i,l}(t), \mu_i(t)) = \int_0^\infty M_{i,l}(t)^T Q M_{i,l}(t) + \Phi\mu_i(t)^2 dt, \tag{49a}$$

s.t.

$$\frac{dM_{i,l}(t)}{dt} = A_i(t) * M_{i,l}(t) + B_i(t)\mu_i(t) + D_i(t). \tag{49b}$$

Eq. (49a) is used to regulate the lateral and angular deviation of the state, and meanwhile consider the lateral control efficiency. Specifically, $M_{i,l}(t)^T Q M_{i,l}(t)$ determines the lateral control efficiency, $\Phi\mu_i(t)^2$ determines the lateral comfort, and $Q = \begin{bmatrix} \pi_1 & \\ & \pi_2 \end{bmatrix}$, $\Phi = \pi_3$ are the pre-define diagonal positive-definite weighting matrices to guarantee the smoothness of the turning angular speed.

By solving the continuous ELQR, the optimal control for the lateral dynamic controller in Eq. (48) is given as a linear feedback and feedforward controller as below:

$$\mu_i(t) = k_{bi}(t)^T M_{i,l}(t) + k_{fi}(t) D_i(t). \tag{49c}$$

where $k_{bi}(t) = [k_{ri}(t), k_{\Delta\theta i}(t)]$ is the continuous feedback gain for the lateral and angular deviation of vehicle $i$ at time $t$, respectively. And $k_{fi}(t)$ is the continuous feedforward gain for the angular change when vehicle proceed. All these gains can be solved by Continuous Algebraic Recatti Equation (Anderson & Moore, 2007).

$$k_{bi}(t) = -R_i^{-1} B_i(t)^T P_i(t), \tag{50a}$$

$$k_{fi}(t) = -R_i^{-1} B_i^T(t) \left[ \left( A_i - R_i^{-1} B_i(t) B_i^T(t) P_i(t) \right)^T \right]^{-1} P_i(t), \tag{50b}$$

$$P_i(t) A_i(t) + A_i^T(t) P_i(t) - P_i(t) B_i(t) R_i^{-1} B_i^T(t) P_i(t) = Q. \tag{50c}$$

### 5.3. *Nonholonomic vehicle lateral dynamics*

To make the system more realistic, a nonholonomic vehicle dynamics system has been considered as an extension. The main difference from the holonomic system is taking the steering angle into account (Necsulescu et al., 2010). As shown in **FIG. 6**, the revised vehicle lateral dynamics state $\mathcal{M}_i(t)$ can be defined as $\mathcal{M}_i(t) = [r_i(t), \Delta\theta_i(t), \delta_i(t)]^T$ where the $\delta_i(t)$ denotes the steering angle of the vehicle, $L_r$ denotes the spacing between the front and rear wheel.

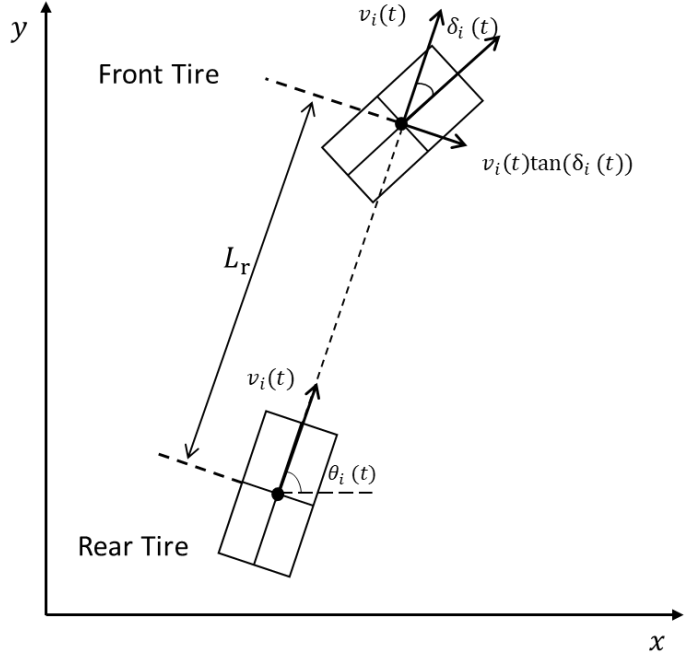

**FIG 6.** Nonholonomic Vehicle Dynamics in Cartesian Coordinates Model

Then, the vehicle dynamics can be described by the state space as below:

$$\frac{d\mathcal{M}_i(t)}{dt} = \begin{pmatrix} v_i(t)\left(1 - \frac{r_i(t)}{R_i(t)}\right)\tan(\Delta\theta_i(t)) \\ \frac{v_i(t)\left(1 - \frac{r_i(t)}{R_i(t)}\right)\tan(\delta_i(t))}{\cos(\Delta\theta_i(t))L_r} - \frac{v_i(t)}{R_i(t)} \\ \kappa_i(t) \end{pmatrix}, \tag{51}$$

where $R_i(t)$ denotes the radius of CAV $i$ at time $t$ for the nonholonomic system. $\kappa_i(t)$, which is the control input, represents the steering angular velocity of CAV $i$. Similar to Eq. (47), we apply a small-angle approximation for $\Delta\theta_i(t)$, which gives:

$$\frac{d\mathcal{M}_i(t)}{dt} = \begin{pmatrix} v_i(t)\left(1 - \frac{r_i(t)}{R_i(t)}\right)\Delta\theta_i(t) \\ \frac{v_i(t)\left(1 - \frac{r_i(t)}{R_i(t)}\right)\tan(\delta_i(t))}{L_r} - \frac{v_i(t)}{R_i(t)} \\ \kappa_i(t) \end{pmatrix}. \tag{52}$$

As can be found that, Eq. (52) still has a non-linear term, $\tan(\delta_i(t))$, where practically $\delta_i(t)$ is not small enough to make the small-angular approximation hold. Thus, to handle the non-linear continuous optimal control problem, a discretization step and a linearization step are usually applied. For discretization, the Forward Euler method (Puwal & Roth, 2007) is adopted, which transfer Eq. (52) to the Eq. (53) below:

$$\mathcal{M}_{i,t+t_s} = y_{i,t} + f(\mathcal{M}_{i,t}, u_{i,t})t_s, \tag{53}$$

where $t_s$ is the time discretization length (i.e. 0.001 sec in our case). Note that the equilibrium of the state space is $\mathcal{M}_e = [0,0,0]^T$, with $u_e = 0$. Then, by the first order Taylor series expansion, we can get the discretized LTV system:

$$\mathcal{M}_{i,t+t_s} \approx \mathcal{M}_{i,t} + \nabla f_{\mathcal{M}}(\mathcal{M},u)t_s|_{\mathcal{M}_{i,t}=\mathcal{M}_e}(\mathcal{M}_{i,t} - \mathcal{M}_e) \tag{54a}$$
$$+ \nabla f_u(\mathcal{M},u)t_s|_{u_{i,t}=u_e}(\kappa_{i,t} - u_e),$$

which can be further simplified as:

$$\mathcal{M}_{i,t+t_s} \approx \left(I + \nabla f(\mathcal{M},u)t_s|_{\mathcal{M}_{i,t}=\mathcal{M}_e}\right)\mathcal{M}_{i,t} + \nabla f_u(\mathcal{M},u)t_s|_{u_{i,t}=u_e} + D_{i,t}\frac{v_{i,t}}{R_i(t)}, \tag{54b}$$

where $\nabla f(\mathcal{M},u)t_s|_{\mathcal{M}_{i,t}=\mathcal{M}_e} = \begin{pmatrix} 0 & v_i(t)t_s & 0 \\ 0 & 0 & \frac{v_i(t)}{L_r}t_s \\ 0 & 0 & 0 \end{pmatrix}$, and $\nabla f_u(\mathcal{M},u)t_s|_{u_{i,t}=u_e} = \begin{pmatrix} 0 \\ 0 \\ t_s \end{pmatrix}$, $D_{i,t} = \begin{pmatrix} 0 \\ -t_s \\ 0 \end{pmatrix}$.

After simplification, the nonlinear system given in Eq. (54b) can be approximated as an LTV system:

$$\mathcal{M}_{i,t+t_s} \approx A^*_{i,t} \times \mathcal{M}_{i,t+t_s} + B^*_{i,t}\kappa_{i,t} + D^*_{i,t}\frac{v_{i,t}}{R_{i,t}}, \tag{55}$$

where $A^*_{i,t} = \begin{pmatrix} 1 & v_i(t)t_s & 0 \\ 0 & 1 & \frac{v_i(t)}{L_r}t_s \\ 0 & 0 & 1 \end{pmatrix}$, $B^*_{i,t} = \begin{pmatrix} 0 \\ 0 \\ t_s \end{pmatrix}$.

By modifying lateral state-space formulation, we can still apply a discrete version of ELQR controller in a rolling horizon fashion.

$$\min J_i(\mathcal{M}_{i,t}, \mu_{i,t}) = \sum_{t=0}^{\infty} \mathcal{M}_{i,t}^T Q \mathcal{M}_{i,t} + R\kappa_{i,t}^2, \tag{56a}$$

s.t.

$$\mathcal{M}_{i,t+t_s} \approx A^*_{i,t} \times \mathcal{M}_{i,t} + B^*_{i,t}\kappa_{i,t} + D^*_{i,t}\frac{v_{i,t}}{R_{i,t}}, \tag{56b}$$

where $Q$ and $R$ are the positive definite weighting matrix for system state and control input. By solving the above discrete ELQR, we can get the control law for the lateral movement as discrete linear feedback and feedforward controller given below:

$$\mathcal{M}_{i,t+t_s} \approx A^*_{i,t} \times \mathcal{M}_{i,t+t_s} + B^*_{i,t}\kappa_{i,t} + D^*_{i,t}\frac{v_{i,t}}{R_{i,t}}, \tag{57a}$$

where $k_{fb}$, and $k_{ff}$ are linear feedback and feedforward gains solved by Discrete Algebraic Recatti Equation (Lancaster & Rodman, 1995) as below:

$$k_{fb} = -\left(R + B_{i,t}^{*T} P_{i,t}^* B_{i,t}^*\right)^{-1} B_{i,t}^{*T} P_{i,t}^* A_{i,t}^*, \tag{57b}$$

$$k_{ff} = -\left(R + B_{i,t}^{*T} P_{i,t}^* B_{i,t}^*\right)^{-1} B_{i,t}^{*T} P_{i,t}^* A_{i,t}^* \left(A_{i,t}^* - P_{i,t}^{*-1}\left(P_{i,t}^* - Q\right)\right)^{-1} D_{i,t}^*, \tag{57c}$$

$$P_{i,t}^* = Q + A_{i,t}^* \left[P_{i,t}^* - P_{i,t}^* B_{i,t}^* \left(R + B_{i,t}^{*T} P_{i,t}^* B_{i,t}^*\right)^{-1} B_{i,t}^{*T} P_{i,t}^*\right] A_{i,t}^*. \tag{57d}$$

**Remark 2.** For a lane change problem or a parallel ramp layout, we can treat the designed trajectory curve as any generic twice-differentiable curve ($\Upsilon$) on a plane by a consecutive $(x, y)$ points in a global Cartesian Coordinate. By the construction of curvilinear coordinate, the differentiable curve can be parametrized as $\Upsilon(s) = \left(p_x(s), p_y(s)\right)$ (Sendra & Winkler, 1991), where $p_x(s)$ and $p_y(s)$ are the global Cartesian Coordinate over the point $s$ on the curve. By the road intelligence, the corresponding curve radius at each point $s$ can be pre-calculated as the reciprocal of the curvature $K(s)$ as below:

$$R(s) = \frac{1}{K(s)} = \frac{1}{\sqrt{\left(\frac{d^2 p_x(s)}{ds^2}\right)^2 + \left(\frac{d^2 p_y(s)}{ds^2}\right)^2}}. \tag{58}$$

During vehicle operation, since the controller is implemented in a rolling horizon, at each time point t, we can update the corresponding value R(t) based on current vehicles position $s$, to implement the ELQR control.

## 6. Numerical experiments

This section presents numerical simulations to validate the efficiency and stability of our proposed control strategy. Both equally weighted and non-equally weighted information cases mentioned above are conducted on MATLAB. Since we design the longitudinal controller as a system-based controller and the lateral controller as a local-based controller, the experimental analysis of these two are discussed separately.

### 6.1. Simulation of longitudinal controller

Firstly, to validate the correctness and effectiveness of the proposed propositions, we set the range of the desired time gap at $\tau = [0.5, 1, 1.5]$ and the number of communicated predecessors $N_i \in [1,6]$ to find the feasible region of control gains for two cases. Based on these settings and the propositions, the results shown in **FIG. 7**. illustrate that the boundaries of the relationship between $\omega_e$ and $\omega_v$ is proportional, and with the number of CAVs in the communication topology ($N_i$) increases, the eligible combination of weight coefficients (the area under the boundaries) increases, which indicates that the longitudinal controller we design is practical. Additionally, this growth trend is reflected in a desired time gap of vehicle, $\tau$. As $\tau$ increases, the area under the boundaries increases. Specifically, though two cases follow the same trends, the equally weighted case shown in **FIG. 7** (a)-(c) has more eligible combinations of the weight coefficients than the non-equal weighted ones shown in **FIG. 7** (d)-(f).

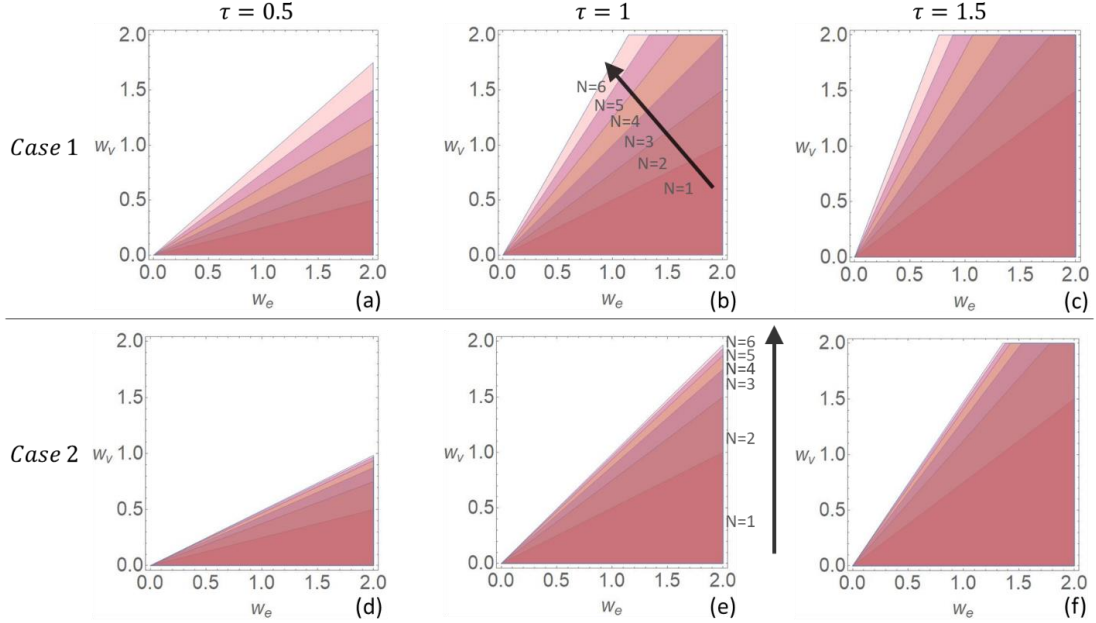

**FIG 7.** Feasible region of control gains for string stability (a)-(c) Equally weighted case, (d)-(f) Non-equally weighted case

Given the control gain analysis, the default parameter values for the CAVs longitudinal controller design are shown in **Table. 1** to conduct CAVs trajectories simulation. The experiment setup consists of 12 CAVs control simulation with one leading vehicle. Since the vehicle initial speed is set to 20 m/s, to reflect our algorithm better, we preset the average distance between two adjacent cars to 20 meters with uniformly distributed $[-10,10]$ meters deviation. Based on that, the origin locations of the CAVs are randomly generated. In this example, the 12 CAVs are initially set up as 7 vehicles on mainline and 5 vehicles on on-ramp whose corresponding relative distance to the merging point are $X_M = [0, -30, -46, -68, -89, -165, -186]$ and $X_R = [-20, -109, -132, -154, -198]$ as shown in **FIG. 8**. By applying the virtual car sequencing method, the initial positions of 12 CAVs on the virtual $Z$-axis are determined as $X_z = [0, -20, -30, -46, -68, -89, -109, -132, -154, -165, -186, -198]$. Meanwhile, the movement of the leading vehicle is generated based on a sine velocity function which indicates the back-and-forth changes of the velocity in a short amount of time that can directly reflect the optimization result of the control algorithm. The mean value of the sine function is set to the initial speed, the oscillation magnitude is set to 3, and the frequency is set to 0.0005 to better visualize the result. The merging point is set to 600 meters away from the first CAV entering the system. To make the simulation closer to the reality, the acceleration threshold is set to $[-3,3] \, m/s^2$. Note that both cases use the exactly identical control parameters mentioned in **Table 1**.

**Table 1.** Default value setting for the longitudinal controller experimental design.

| Parameters | Value |
| --- | --- |
| Time Discretization Rate $t_s$ | 0.001 |
| Total Running Time $T$ | 80 sec |
| $\tau$ | 1 sec |
| $L$ | 5 m |
| $v_{initial}$ | 20 m/s |
| $\omega_e$ | 1.4 |

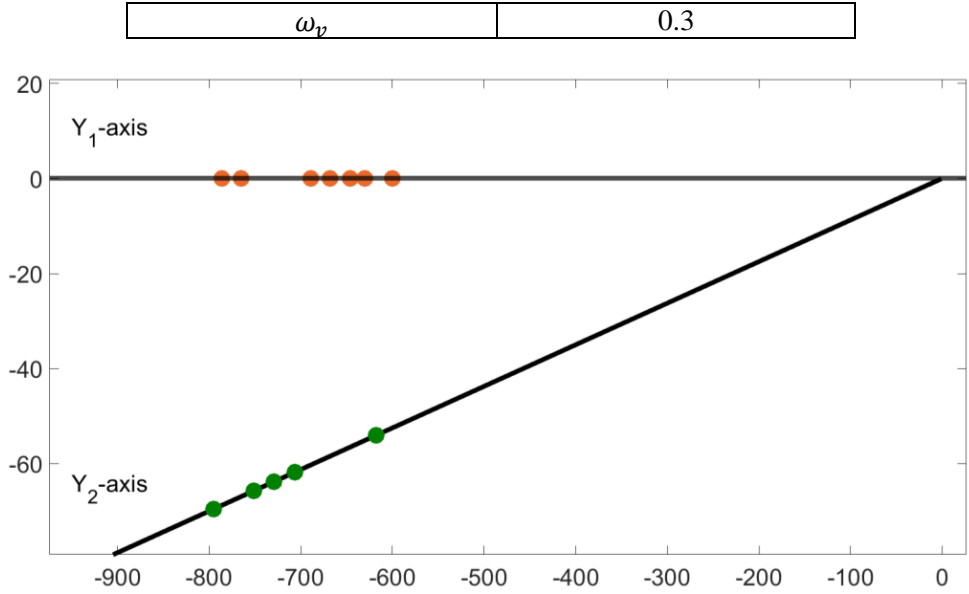

| $\omega_v$ | 0.3 |

**FIG 8**. The initial position of 12 CAVs on $y_1$ and $y_2$ axis.

**FIG. 9** below shows the state evolution of a) position of the 12 CAVs on virtual Z-axis, b) the speed, c) the acceleration and d) the position of CAVs on the mainline ($Y_1$-axis) for the equally weighted information case. Similarly, **FIG. 10** shows the results for Case 2, which is a non-equally weighted information case. It is observed that the disturbances are dampened along the string of CAVs in both state of speed and acceleration. Additionally, the state of position for CAVs on virtual Z-axis and real-world mainline $Y_1$-axis are well reflected the shrinking trend. Thus, this trend illustrates that not only the CAVs on the mainline and on the on-ramp are string stable, but also the mapped virtual platoon on the virtual Z-axis is string stable. Meanwhile, the state of the position results also demonstrates that CAVs are controlled stably in the set headway to ensure safety. Further, the result validates the proposed controller can actively reduce the voids in a small amount of time and meanwhile guarantee the string stability of the whole system in both cases.

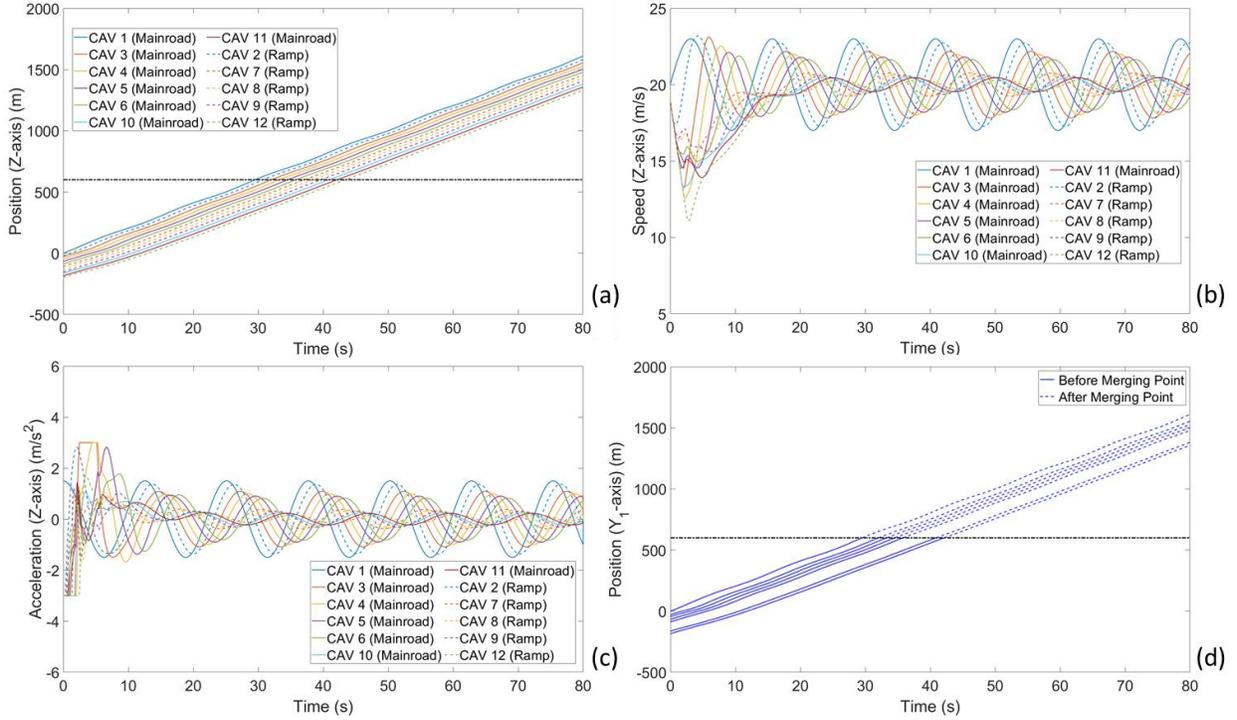

**FIG 9.** State evolution results of longitudinal controller for Case 1: (a) platoon position on $Z$-axis (b) speed (c) acceleration (d) platoon position on $Y_1$-axis

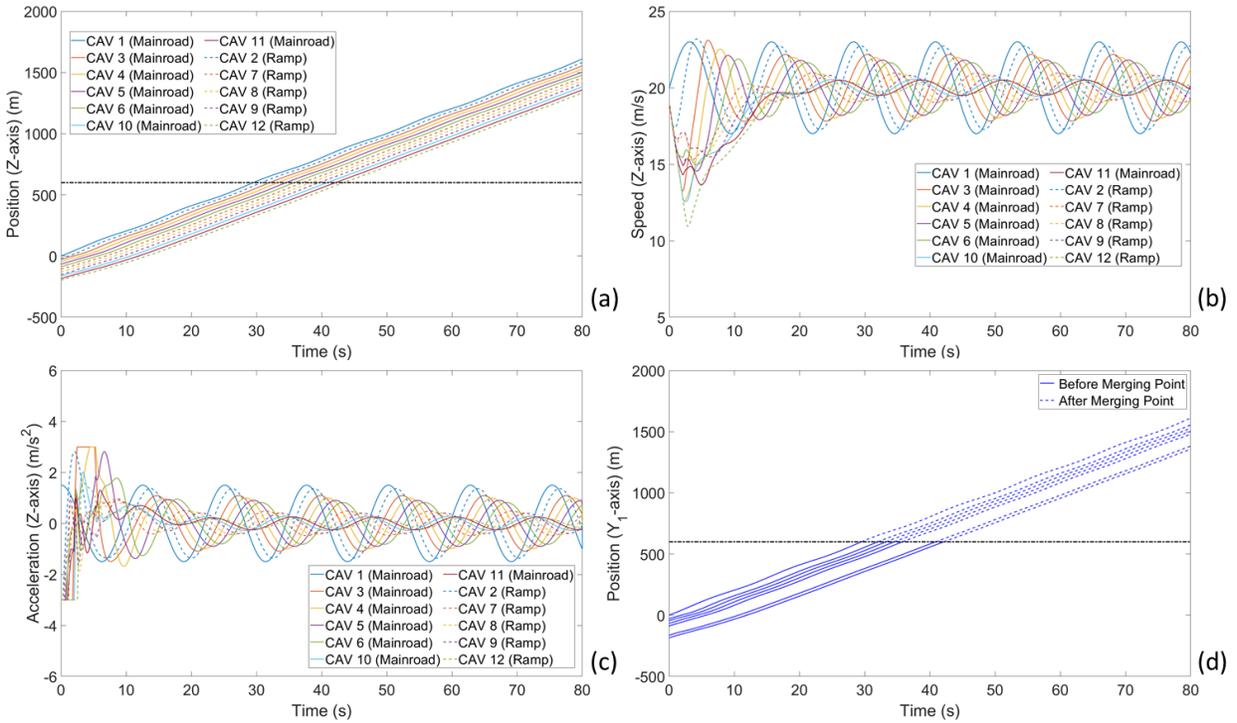

**Fig 10.** State evolution results of longitudinal controller for Case 2: (a) platoon position on $Z$-axis (b) speed (c) acceleration (d) platoon position on $Y_1$-axis

To more quantitatively validate the proposed longitudinal controller that satisfies the propositions mentioned in section 4, we use the square of $L_2$ norm of the absolute velocity for each CAV to verify. **FIG. 11** shows the result of the square of $L_2$ norm of the absolute velocity (energy) for cases 1 and 2. The trends for both cases elucidate that the energy is attenuated through the control of vehicles. Specifically, this attenuation trend applies to CAVs on $Y_1$-axis (red bars), on $Y_2$-axis (blue bars), and on the virtual $Z$-axis. Thus, the results validate the string stability criterion we proposed in Section 4.

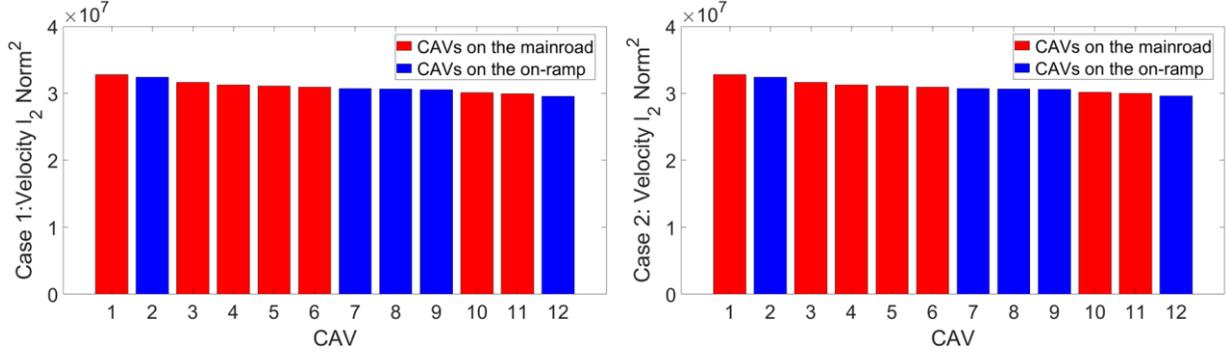

**FIG 11.** Square of $l_2$ norm of absolute velocity comparison for case 1 and case 2.

### 6.2. The test of convergence range to equilibrium state

A test of convergence range to equilibrium state has been designed for the proposed longitudinal controller to examine whether the merging control range is ample enough before merging. Specifically, an extreme case has been conducted to find the minimum control range that allows CAVs on the virtual axle approach to equilibrium state before merging. The extreme case is defined by considering the safety constraint and emerge deceleration behavior. More detailed, this case consists of 4 CAVs, 2 on mainline and 2 on on-ramp, where the initial positions of the CAVs on the virtual $Z$-axis violate the safety constraint (smaller or equal to the headway). Meanwhile, the leading vehicle is designed to brake sharply in a short amount of time during the simulation to test the resilience of the proposed controller. For consistency, the experiment follows the default settings in **Table 1** but with 40 seconds simulation time. The initial positions of CAVs on the mainline are $X_M = [0, -8]$, and the initial positions of CAVs on the on-ramp are $X_R = [-2, -9]$. By applying the virtual car sequencing method, the initial position of 4-CAV platoon on the virtual axle are $X_Z = [0, -2, -8, -9]$. The leading vehicle is set to decelerate at $2\ ms^{-2}$ after running for 13 sec until the velocity reach $10 ms^{-1}$, then accelerate again at $2\ ms^{-2}$ after running for 26 sec until reach the initial velocity, $20 ms^{-1}$. Note that, since the mainline/ramp indicator set $\tilde{F}_z = [1,0,1,0]$, the weighted coefficients are equal in both equally weighted and non-equally weighted case. Thus, we only apply the equally weighted case string stability criterion in this analysis. **FIG. 12** illustrates the 4 CAVs' state evolution of a) position on $Z$-axis, b) deviation of the target spacing, and c) the square of $l_2$ norm of absolute velocity of the 4 CAVs.

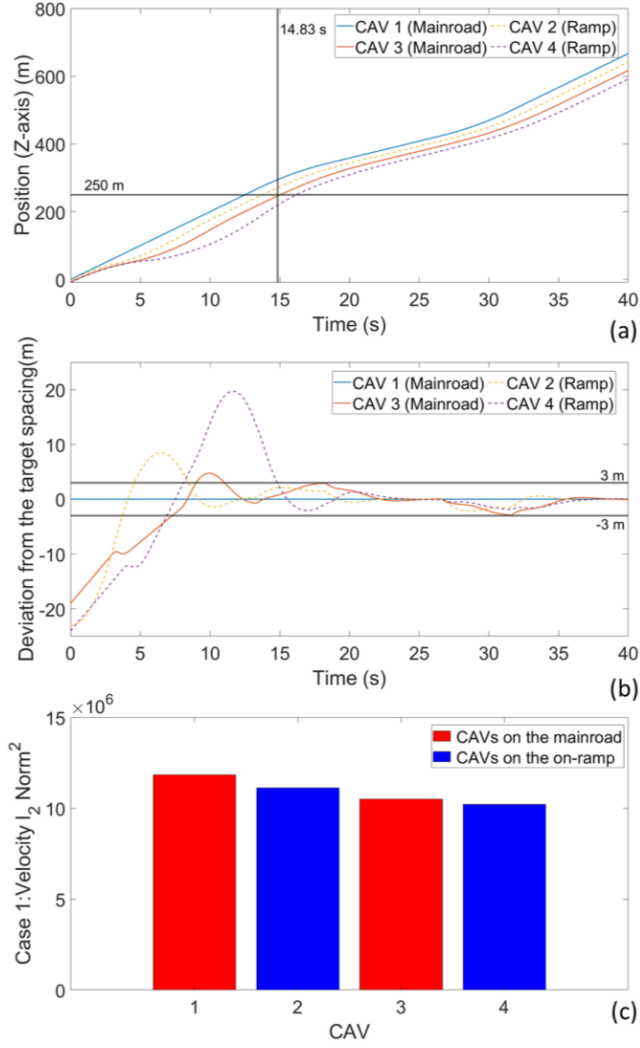

**FIG 12.** Test of convergence range result: (a) platoon position on $Z$-axis (b) deviation from the target spacing (c) square of $l_2$ norm of absolute velocity.

Deviation from target spacing (i.e., $x_{i-1}(t) - x_i(t) - d_{i,1}(t)$) has been used to verify whether the CAVs reach equilibrium close enough for merging CAVs. Specifically, in this case, we define that when the deviation from target spacing meets the range $[-3,3]$, the corresponding CAV gets close to equilibrium and can merge to the mainline relatively safely. In **FIG. 12** (b), it is observed that, until the fourth CAV drive for 14.83 sec, the whole platoon on the virtual axle meets the range $[-3,3]$ which close enough to the equilibrium state and the corresponding travelling distance is around 220 m. This distance is shorter than the length of most merging system, which reveal the applicability of the proposed scheme. Thus, we can conclude that the assumption we made in Section 2 is ample enough for the proposed controller.

*6.3.   Simulation of lateral controller*

Next, we evaluate the extended lateral controller by another numerical experiment. Since we design the lateral controller as a local controller, each CAV is controlled individually, the experiment is designed to focus on one of the CAVs in the platoon. As mentioned in section 5, the lateral movement is more related to individual vehicle stability. Thus, we determine the state evolution of lateral deviation and angular deviation to validate the effectiveness of the lateral controller. The parameter settings are given in **Table 2** as an example. Note that the initial angular deviation is set to 10° to satisfy the small-angle approximation. In this experiment, we select #2 CAV in the platoon, the first CAV entering the on-ramp lane, as the test subject. Since the initial position of the test CAV is located at -620 meter on $Y_2$-axis, the $\theta_{des}$ can be calculated as $36°$. **FIG. 13** demonstrates the schematic diagram of the lateral experiment in a real-world scene.

**Table 2.** Default value setting for the lateral controller experimental design.

| Parameters | Value |
|---|---|
| Initial lateral deviation | 0.4 |
| Initial angular deviation | -10° |
| Radius of the on-ramp | 1000 m |
| T | 80 sec |
| $L_r$ | 2 m |

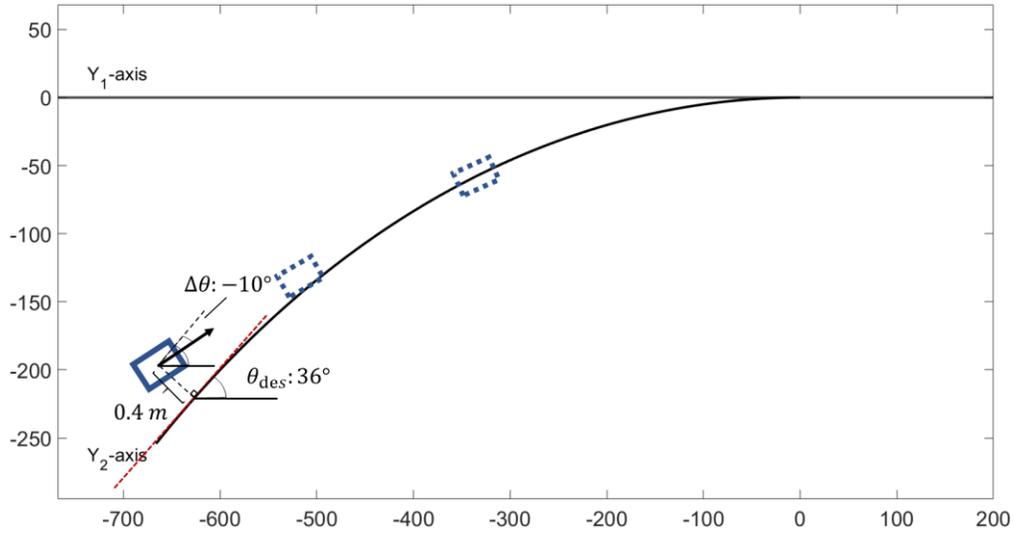

**FIG 13.** Schematic diagram of lateral experiment in real-world scene

The evaluation of the holonomic system is based on two results: 1) the state evolution of lateral deviation and 2) angular deviation for the first on-ramp vehicle as shown in **FIG. 14** (a), (b), since other on-ramp vehicles also exhibit the same conclusion. The result illustrates that the CAV controlled by the lateral controller converges to the road's centerline (infinitely close to 0) very quickly. However, it is observed that, at the time 32 sec, both deviations have a slight peak of changes (disturbances) which indicates that the CAV is passing through the merging point. Due to the small amount of change and instant amount of time, the changes can be neglected.

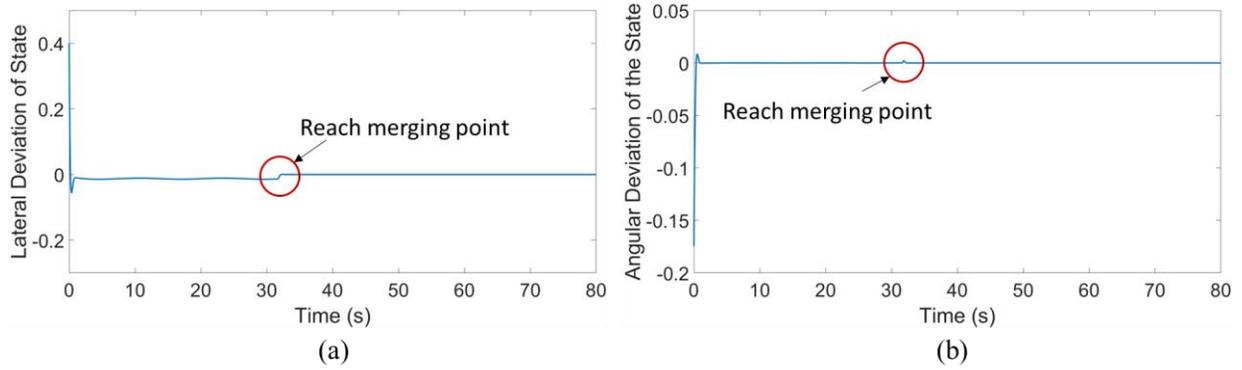

**FIG 14.** Holonomic system results: state evolution of (a) lateral deviation and (b) angular deviation.

The evaluation of the nonholonomic lateral controller has a corresponding steering angle validation term besides lateral and angular deviation. The initial steering angle of the testing CAV is set to 0. Similar trends have been observed in the results shown in **FIG. 15**. Not only the lateral and angular deviation converge fast, but the vehicle steering angle also quickly converges to the designed road steering angle at the same time. Note that, after the vehicle merging to the mainline, the steering angle continuously to be infinitely close to 0 indicates that the vehicle is driving along a straight line. Overall, both holonomic and nonholonomic results well-reflect the control efficiency and smoothness of the lateral movement.

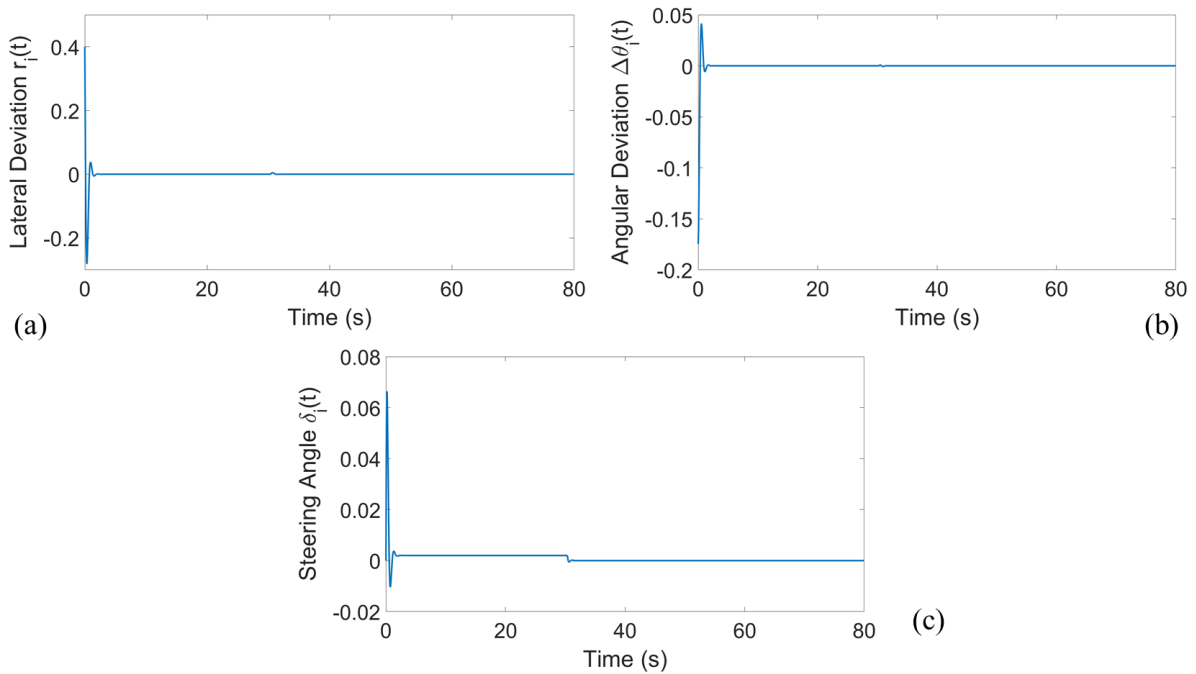

**FIG 15.** Nonholonomic system results: state evolution of (a) lateral deviation, (b) angular deviation, and (c) steering angle.

## 7. Conclusion

This paper proposes a novel cooperative merging control strategy of CAVs based on a virtual rotation approach. Specifically, the virtual rotation strategy has been introduced which project CAVs on mainline and CAVs on on-ramp to a shared straight axis, which greatly simplified the problem to a virtual car following problem while maintaining all characteristic of concern. By using the merge point as a reference, a virtual car following sequence of vehicles is generated by integrating the virtual rotation and FIFO rules, which serves as an upper-level controller. A unidirectional multi-leader communication topology has been specifically designed to fully utilize the downstream CAVs information and achieve strict string stability for upstream mainline and on-ramp vehicles, respectively. With the proposed topology, we design a multiple-predecessor feedback and feedforward longitudinal controller and derived sufficient conditions for the string stability criterion with two case studies: 1) Equally weighted predecessors information case, 2) Non-equally weighted predecessors information case. Further, to expand the application scenarios, we design a local-based lateral controller based on an extended linear-quadratic regulator in a curvilinear coordinate system for the curvy ramp lane case. Two sets of numerical simulation experiments are conducted to verify the efficiency and stability for the system. For longitudinal controller experiments, 12-CAV trajectories portfolio including position, speed, acceleration, and energy have been conducted to validate the model. The results reveal that the proposed controller can actively reduce the voids and meanwhile guarantee the distribution attenuation of the whole system. Similarly, for the lateral control experiment, the result elucidates reasonably well to guarantee the smoothness of the turning movements with the small-angle approximation.

To the best of our acknowledgment, it is the first attempt using the virtual rotation concept to turn a two-dimensional merging problem into a one-dimensional CF control problem for multi-vehicle controller with mathematical proofs of string stability criterion. The result of this study not only proposed a well-defined control strategy ensures safety and smooth the traffic dynamics, but also performs rigorous mathematical modeling of the problem to theoretically illustrate properties.

Some future works are desired on the results of this study. For example, future work could consider systematic car sequencing optimization to further improve the performance in this paper. In addition, future work will extend the current framework to a lane change and parallel ramp layout in a more systematic way.

**Acknowledgements**

This study is partially supported by National Key R&D Program of China (2019YFB1600100), NSFC (71901038), Natural Science Basic Research Plan in Shaanxi Province (2020JQ-392), and Research Funds for the Central Universities, Chang'an University (300102240301), and fully supported by the Wisconsin Traffic Operations and Safety (TOPS) Laboratory.

**Author Contribution Statement**

**Tianyi Chen:** Conceptualization, Methodology, Experiment, Analysis, Writing-original Draft Preparation. **Meng Wang:** Methodology, Writing-Reivew and Editing. **Siyuan Gong:** Experiment, Writing-Review and Editing. **Yang Zhou:** Conceptualization, Methodology, Writing-Review and Editing, Supervision. **Bin Ran:** Funding Support, Writing-Review and Editing.

**References**

Ahn, S, & Cassidy, M. J. (2007). Freeway traffic oscillations and vehicle lane-change maneuvers. *Proceedings of the 17th International Symposium on Traffic and Transportation Theory*.

Ahn, S, Laval, J., & Cassidy, M. J. (2010). Effects of merging and diverging on freeway traffic oscillations. *Transportation Research Record*, *2188*, 1–8. https://doi.org/10.3141/2188-01

Akpakwu, G. A., Silva, B. J., Hancke, G. P., & Abu-Mahfouz, A. M. (2017). A Survey on 5G Networks for the Internet of Things: Communication Technologies and Challenges. In *IEEE Access*. https://doi.org/10.1109/ACCESS.2017.2779844

Anderson, B. D. O., & Moore, J. B. (2007). *Optimal control: linear quadratic methods*. Courier Corporation.

Arfken, G. B., Weber, H. J., & Harris, F. E. (2013). Mathematical Methods for Physicists. In *Mathematical Methods for Physicists*. https://doi.org/10.1016/C2009-0-30629-7

Bian, Y., Zheng, Y., Ren, W., Li, S. E., Wang, J., & Li, K. (2019). Reducing time headway for platooning of connected vehicles via V2V communication. *Transportation Research Part C: Emerging Technologies*, *102*(March), 87–105. https://doi.org/10.1016/j.trc.2019.03.002

Cao, W., Mukai, M., Kawabe, T., Nishira, H., & Fujiki, N. (2015). Cooperative vehicle path generation during merging using model predictive control with real-time optimization. *Control Engineering Practice*. https://doi.org/10.1016/j.conengprac.2014.10.005

Carlson, R. C., Papamichail, I., Papageorgiou, M., & Messmer, A. (2010a). Optimal mainstream traffic flow control of large-scale motorway networks. *Transportation Research Part C: Emerging Technologies*, *18*(2), 193–212. https://doi.org/10.1016/j.trc.2009.05.014

Carlson, R. C., Papamichail, I., Papageorgiou, M., & Messmer, A. (2010b). Optimal motorway traffic flow control involving variable speed limits and ramp metering. *Transportation Science*, *44*(2), 238–253. https://doi.org/10.1287/trsc.1090.0314

Cassidy, M. J., & Rudjanakanoknad, J. (2005). Increasing the capacity of an isolated merge by metering its on-ramp. *Transportation Research Part B: Methodological*, *39*(10), 896–913. https://doi.org/10.1016/j.trb.2004.12.001

Chen, D., & Ahn, S. (2018). Capacity-drop at extended bottlenecks: Merge, diverge, and weave. *Transportation Research Part B: Methodological*, *108*, 1–20. https://doi.org/10.1016/j.trb.2017.12.006

Chen, N., van Arem, B., Alkim, T., & Wang, M. (2020). A Hierarchical Model-Based Optimization Control Approach for Cooperative Merging by Connected Automated Vehicles. *IEEE Transactions on Intelligent Transportation Systems*, 1–14. https://doi.org/10.1109/tits.2020.3007647

Darbha, S., Konduri, S., & Pagilla, P. R. (2017). Effects of V2V communication on time headway for autonomous vehicles. *Proceedings of the American Control Conference*. https://doi.org/10.23919/ACC.2017.7963246

Darbha, S.Konduri, S., & Pagilla, P. R. (2019). Benefits of V2V communication for autonomous and connected vehicles. *IEEE Transactions on Intelligent Transportation Systems*. https://doi.org/10.1109/TITS.2018.2859765